\documentclass[modern]{aastex62}

\usepackage{graphicx}
\usepackage{amsbsy}
\usepackage{amsmath}
\usepackage{epsf}
\usepackage{paralist}
\usepackage{tablefootnote}
\usepackage{geometry}
\usepackage{wrapfig}

\newcommand{\beq}{\begin{equation}}
\newcommand{\eeq}{\end{equation}}

\DeclareMathAlphabet{\mathsfsl}{OT1}{cmss}{bx}{sl}
\SetMathAlphabet{\mathsfsl}{bold}{OT1}{cmss}{bx}{sl}

\usepackage{amssymb}
\usepackage{pifont}

\usepackage{hyperref}

\usepackage{epstopdf}
\begin{document}

\title{PASIPHAE: A high-Galactic-latitude, high-accuracy optopolarimetric survey}



\author{Konstantinos Tassis}
\affiliation{Department of Physics and ITCP, University of Crete, Voutes, 70013 Heraklion, Greece}
\affiliation{Institute of Astrophysics, Foundation for Research and Technology-Hellas, Voutes, 70013 Heraklion, Greece}

\author{Anamparambu N. Ramaprakash}
\affiliation{Inter-University Centre for Astronomy and Astrophysics, Post Bag 4, Ganeshkhind, Pune - 411 007, India}

\author{Anthony C. S. Readhead}
\affiliation{Cahill Center for Astronomy and Astrophysics, California Institute of Technology, 1200 E California Blvd, MC 249-17, Pasadena CA, 91125, USA } 

\author{Stephen B. Potter}
\affiliation{South African Astronomical Observatory, PO Box 9, Observatory, 7935, Cape Town, South Africa}

\author{Ingunn K. Wehus}
\affiliation{Institute of Theoretical Astrophysics, University of Oslo, P.O. Box 1029 Blindern, NO-0315 Oslo, Norway}

\author{Georgia V. Panopoulou}
\affiliation{Cahill Center for Astronomy and Astrophysics, California Institute of Technology, 1200 E California Blvd, MC 249-17, Pasadena CA, 91125, USA }

\author{Dmitry Blinov}
\affiliation{Department of Physics and ITCP, University of Crete, Voutes, 70013 Heraklion, Greece}
\affiliation{Institute of Astrophysics, Foundation for Research and Technology-Hellas, Voutes, 70013 Heraklion, Greece}

\author{Hans Kristian Eriksen}
\affiliation{Institute of Theoretical Astrophysics, University of Oslo, P.O. Box 1029 Blindern, NO-0315 Oslo, Norway}

\author{Brandon Hensley}
\affiliation{Department of Astrophysical Sciences,  Princeton University, Princeton, NJ 08544, USA} 

\author{Ata Karakci}
\affiliation{Institute of Theoretical Astrophysics, University of Oslo, P.O. Box 1029 Blindern, NO-0315 Oslo, Norway}

\author{John A. Kypriotakis}
\affiliation{Department of Physics and ITCP, University of Crete, Voutes, 70013 Heraklion, Greece}
\affiliation{Institute of Astrophysics, Foundation for Research and Technology-Hellas, Voutes, 70013 Heraklion, Greece}

\author{Siddharth Maharana}
\affiliation{Inter-University Centre for Astronomy and Astrophysics, Post Bag 4, Ganeshkhind, Pune - 411 007, India}

\author{Evangelia Ntormousi}
\affiliation{Institute of Astrophysics, Foundation for Research and Technology-Hellas, Voutes, 70013 Heraklion, Greece}

\author{Vasiliki Pavlidou}
\affiliation{Department of Physics and ITCP, University of Crete, Voutes, 70013 Heraklion, Greece}
\affiliation{Institute of Astrophysics, Foundation for Research and Technology-Hellas, Voutes, 70013 Heraklion, Greece}

\author{Timothy J. Pearson}
\affiliation{Cahill Center for Astronomy and Astrophysics, California Institute of Technology, 1200 E California Blvd, MC 249-17, Pasadena CA, 91125, USA }

\author{Raphael Skalidis}
\affiliation{Department of Physics and ITCP, University of Crete, Voutes, 70013 Heraklion, Greece}
\affiliation{Institute of Astrophysics, Foundation for Research and Technology-Hellas, Voutes, 70013 Heraklion, Greece}


\begin{abstract}

PASIPHAE (the Polar-Areas Stellar Imaging in Polarization High-Accuracy Experiment) is an optopolarimetric survey aiming to measure the linear
polarization from millions of stars, and use these to create a
three-dimensional tomographic map of the magnetic field threading dust clouds within the Milky Way. This map will provide invaluable information for future CMB B-mode
experiments searching for inflationary gravitational waves, providing
unique information regarding line-of-sight integration effects.
Optical polarization observations of a large number of stars at known distances, tracing the same dust that emits polarized microwaves, can map the magnetic field between them. The Gaia mission is measuring distances to a billion stars, providing an opportunity to produce a tomographic map of Galactic magnetic field directions, using optical polarization of starlight. Such a map will not only boost CMB polarization foreground removal, but it will also have a profound impact in a wide range of astrophysical research, including interstellar medium physics, high-energy astrophysics, and evolution of the Galaxy. Taking advantage of the novel technology implemented in our high-accuracy Wide-Area Linear Optical Polarimeters (WALOPs) currently under construction at IUCAA, India, we will engage in a large-scale optopolarimetric program that can meet this challenge: a survey of both northern and southern Galactic polar regions targeted by CMB experiments, covering over 10,000 square degrees, which will measure linear optical polarization of over 360 stars per square degree (over 3.5 million stars, a 1000-fold increase over the state of the art) and deliver at least a $3 \sigma$ measurement for individual stars with polarization fraction $p \approx 0.5\%$. The survey will be conducted concurrently from the South African Astronomical Observatory in Sutherland, South Africa in the southern hemisphere, and the Skinakas Observatory in Crete, Greece, in the north.

\end{abstract}


\section{Introduction}\label{intro}
\subsection{Galactic dust a major obstacle in the hunt for CMB B-modes}\label{intro1}

\begin{wrapfigure}{R}{0.5\textwidth}
\centering
\includegraphics[width=0.49\textwidth, clip]{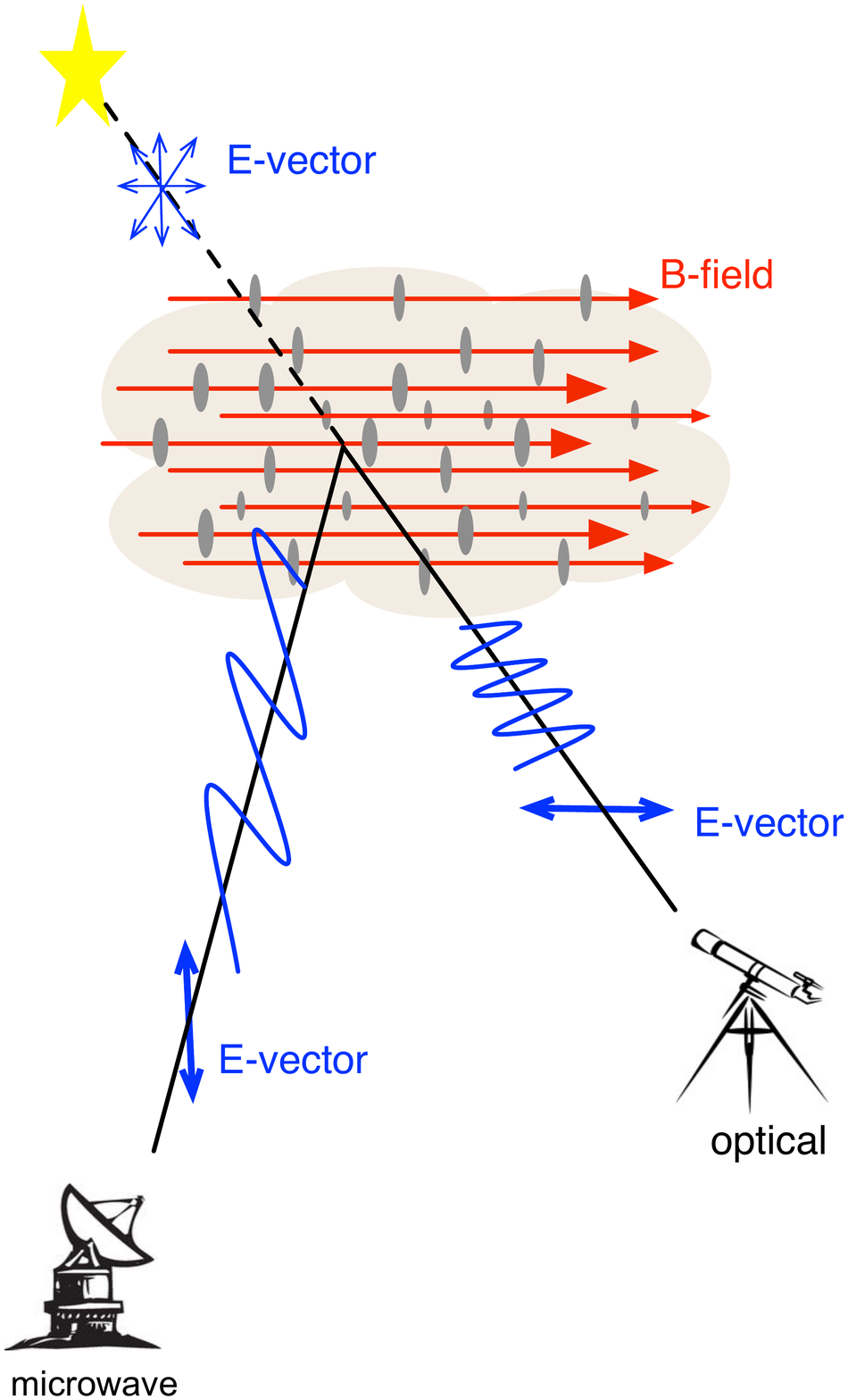}
\caption{Polarization of dust emission in the microwaves and of starlight at optical wavelengths (due to dichroic dust absorption); both result from the alignment of the dust grain short axis with the direction of the magnetic field. 
\label{fig:cartoon1}}
\end{wrapfigure}

The polarization of the cosmic microwave background (CMB) is hailed as a paramount messenger from the early universe. Primordial gravitational waves produced by cosmic inflation are predicted to impart a characteristic B-mode pattern to the CMB polarization \citep{1985SvA....29..607P, 1997PhRvD..55.7368K, 1997NewA....2..323H}. Detecting such a signal would constitute one of the most important discoveries in physics of all time. It would reveal the physical mechanism responsible for the Big Bang itself, one of the greatest mysteries ever; it would be our first experimental probe of yet-uncharted physics territory, including the physics of the universe during an initial de Sitter phase of exponential expansion, when quantum gravity must have seeded all the structure in the universe that we see today.
For this reason, such a signal is aggressively pursued by experiments around the world, including current, planned, and proposed ground-based, suborbital and space observatories such as BICEP3\footnote{\url{https://www.cfa.harvard.edu/CMB/bicep3/}}, AdvACT\footnote{\url{http://act.princeton.edu/overview/camera-specifications}}, Spider\footnote{\url{https://spider.princeton.edu/}}, POLARBEAR-2\footnote{\url{https://arxiv.org/abs/1608.03025}}, CMB-S4\footnote{\url{https://cmb-s4.org/}}, PIPER\footnote{\url{https://asd.gsfc.nasa.gov/piper/}}, LiteBIRD\footnote{\url{http://litebird.jp/eng/}}, PRISM\footnote{\url{http://www.prism-mission.org/}}, Simons Observatory\footnote{\url{https://simonsobservatory.org/}}, and others.

In 2014, a claim by the BICEP2 team of the detection of a B-mode signal \citep{2014PhRvL.112x1101B} in the region of the sky around the southern Galactic pole rallied much enthusiasm in the physics community. However, follow-up studies by Planck \citep{2015PhRvL.114j1301B} demonstrated that the BICEP2 signal was contaminated by the ever-present polarized emission from cold interstellar dust within our Galaxy.

Galactic dust is cold and it emits substantial amounts of optically thin, thermal radiation at the CMB frequencies. This emission is polarized (see Fig. \ref{fig:cartoon1}, ``microwave"), because dust grains tend to align their short axis with the interstellar magnetic field [for a recent review on grain alignment, see \citet{2015ARA&A..53..501A}]. Grains in turn radiate with the E-vector of the radiation preferentially polarized along their long axis. The polarization of grain emission has a B-mode component that can mimic a primordial signal \citep{2014PhRvL.112x1101B, 2015PhRvL.114j1301B}.

To overcome contamination from polarized dust emission, experiments chasing after CMB B-modes implement a two-fold approach. First, they target regions of the sky away from the Galactic plane that are ``cleanest" of dust. Second, they subtract the dust contribution from maps of the polarized microwave sky using empirical models, which are optimized in the following way: dust is mapped at frequencies higher than the CMB, where its thermal emission dominates; its polarization pattern is then extrapolated to lower, CMB-dominated frequencies \citep{2016A&A...586A.133P}.

However, Planck has now demonstrated that this approach is not adequate, for two reasons. First, because no region of the sky is ``clear enough": a dusty veil obscures our view of the early universe {\em wherever we look} \citep{2014PhRvL.112x1101B, 2015PhRvL.114j1301B}. Second, because 3-dimensional effects might decorrelate the polarization maps between different frequencies, reducing the accuracy of empirical extrapolation models below the level required to detect a primordial B-mode signal \citep{2015MNRAS.451L..90T, 2017A&A...599A..51P, 2018arXiv180104945P}.

\subsection{3D structure of dust clouds: an irreducible noise floor?}\label{intro2}

\begin{wrapfigure}{R}{0.5\textwidth}
\centering
\includegraphics[width=0.4\textwidth, clip]{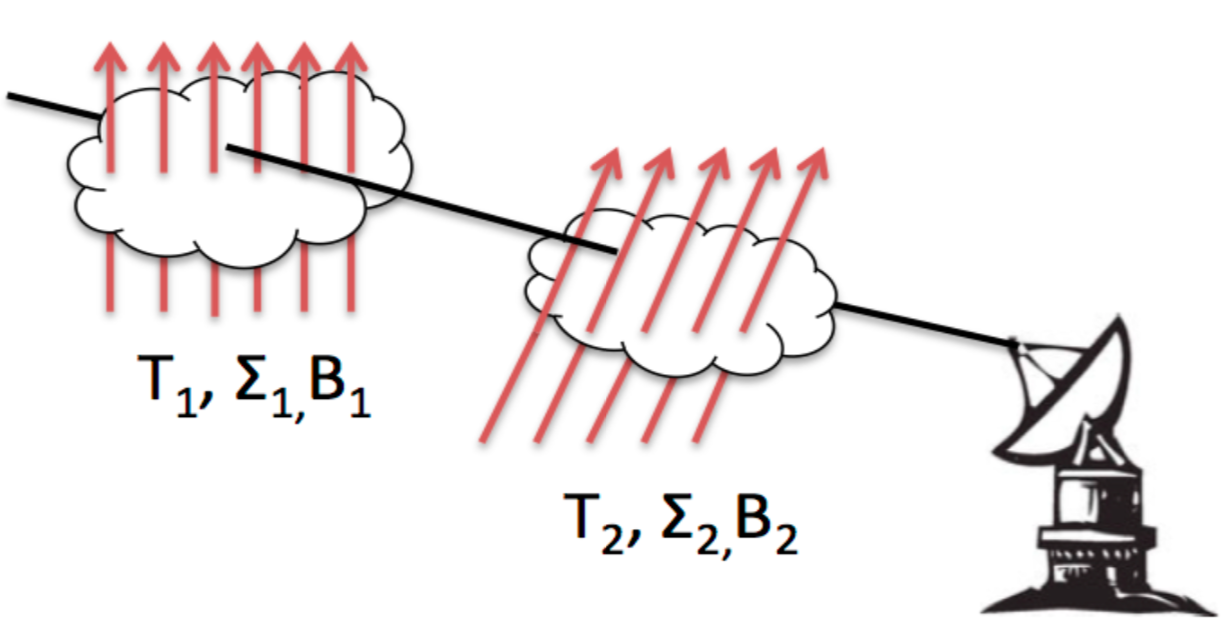}
\caption{Two clouds with different temperatures T, dust column densities $\Sigma$, and magnetic fields $\vec{B}$ contribute to the total dust-emission intensity along a line of sight \citep{2015MNRAS.451L..90T}.
\label{fig:cartoon2}}
\end{wrapfigure}

Microwave maps record the {\em integrated} emission from all dust clouds along the line of sight. However, this integrated emission is the sum of contributions from different clouds with different physical properties (Fig. \ref{fig:cartoon2}) and different thermal emission intensities at different frequencies. If the magnetic fields in different clouds are misaligned, the change in relative contribution from each cloud will lead to shifts of the polarization pattern between frequencies. In the most extreme scenario, one can imagine one cloud dominating at the dust-mapping frequency, and another cloud (with a completely different magnetic field orientation, and thus a completely different polarization pattern) dominating at the CMB frequency. This effect was first identified by \citet{2015MNRAS.451L..90T} and its expected level was explored by a series of studies [e.g., \citet{2016A&A...586A.133P, 2017A&A...599A..51P, 2018arXiv180104945P, 2017PhRvD..95j3511P, 2017MNRAS.469.2982P, 2018MNRAS.476.1310M, 2018ApJ...853..127H}].
Because this ``noise level" is set by nature (the 3D distribution of interstellar dust and magnetic fields in the Milky Way), it constitutes an irreducible noise floor, unless a way is identified to experimentally map this distribution.

To astrophysicists, interstellar dust and interstellar magnetic fields have long been known to lie at the heart of some of the most fundamental open problems in our understanding of the visible Universe. For interstellar medium physicists, they constitute key ingredients of interstellar clouds, and they both control and reveal how interstellar clouds form, fragment, and collapse to one or multiple stars. By providing a unique view on how stars form in cosmic structures dominated by dark matter, they connect the dark sector with the physics of the observable universe. For high-energy particle astrophysics, the Galactic magnetic field is a ubiquitous distorter of the trajectories of relativistic particles, making it impossible to directly track these particles back to their sources. Moreover, the BICEP2 and Planck results have made interstellar dust a topic of critical importance for every physicist, as it is now clear that new physics, as recorded in photons from the early Universe, has to be viewed through our own Galaxy's dusty veil. Our need to understand interstellar dust and its polarization properties has thus gained a new urgency.

{\em Our path to inflationary B-modes passes through developing the technology to perform interstellar medium tomography, and this is a challenge that the interstellar medium community will have to meet if we are to take the next leap forward in fundamental physics and cosmology}.  This is the scope of our project.

PASIPHAE aims to perform a survey of unprecedented scale of the northern and southern sky in optical polarization, which traces interstellar dust and the magnetic field within which it lies. Our survey, in combination with stellar distances from the on-going ESA's Gaia mission, will usher in a new era of 3-dimensional, tomographic cartography of the Galactic magnetic field, leading to breakthroughs in foreground control during the search for CMB B-modes, as well as in the fields of interstellar medium astrophysics, star formation, gamma-ray astronomy, and cosmic-ray physics. All data collected from this major instrumentation and observational effort will be made publicly available for use by the scientific community.

\subsection{A new era begins: magnetic tomography of the Milky Way}\label{tomography}
 
Most often, starlight is inherently unpolarized when it leaves its parent star. When it passes through a dust cloud, however, some of it is absorbed, and the fraction that emerges is partially linearly polarized. This polarization is a complementary effect of the same astrophysical process that causes the polarization of dust emission (the major B-mode foreground). Dust absorption preferentially attenuates the component of the electric field of starlight along the long axis of the aspherical dust grains. The light that makes it through is thus partially polarized perpendicularly to this axis (along the magnetic field, see Fig. \ref{fig:cartoon1} ``optical", black solid line). Emission and absorption by the same population of dust grains \citep{2008ApJ...674..304W, 2008MNRAS.387..797H} thus traces the same magnetic field, and the polarizations of emitted and transmitted light will be orthogonal to each other \citep{2015A&A...576A.106P}.
If we knew the optical polarization properties of a large number of stars in the same small region of sky, and these stars were located at different, known distances, we could deduce the 3-dimensional structure of the interstellar medium (clouds, amount of dust, magnetic fields) in between these stars; tomographic cartography of dust clouds, necessary to clear the path to the primordial B-modes, would be possible.

The ongoing Gaia mission of ESA\footnote{\url{http://sci.esa.int/gaia/}} is in the process of measuring extremely accurately distances (from parallaxes) for a billion stars over the entire sky. This program will yield more than enough stellar distances to perform tomography at the needed resolution, and it presents a unique opportunity and challenge -- to provide a similar leap in the number of stars with measured optical polarization properties.

PASIPHAE is such a program, which will enhance our CMB B-mode foreground subtraction capabilities, and at the same time strongly complement, and add very significant to, all currently running or planned CMB foreground mapping programs at microwave and radio frequencies.

\subsection{Optopolarimetry away from the Galactic plane: the requirements for state of the art magnetic field tomography} \label{survey_req}

\subsubsection{What part of the sky do we need to observe?}

\begin{wrapfigure}{L}{0.6\textwidth}
\centering
\includegraphics[width=0.58\textwidth, clip]{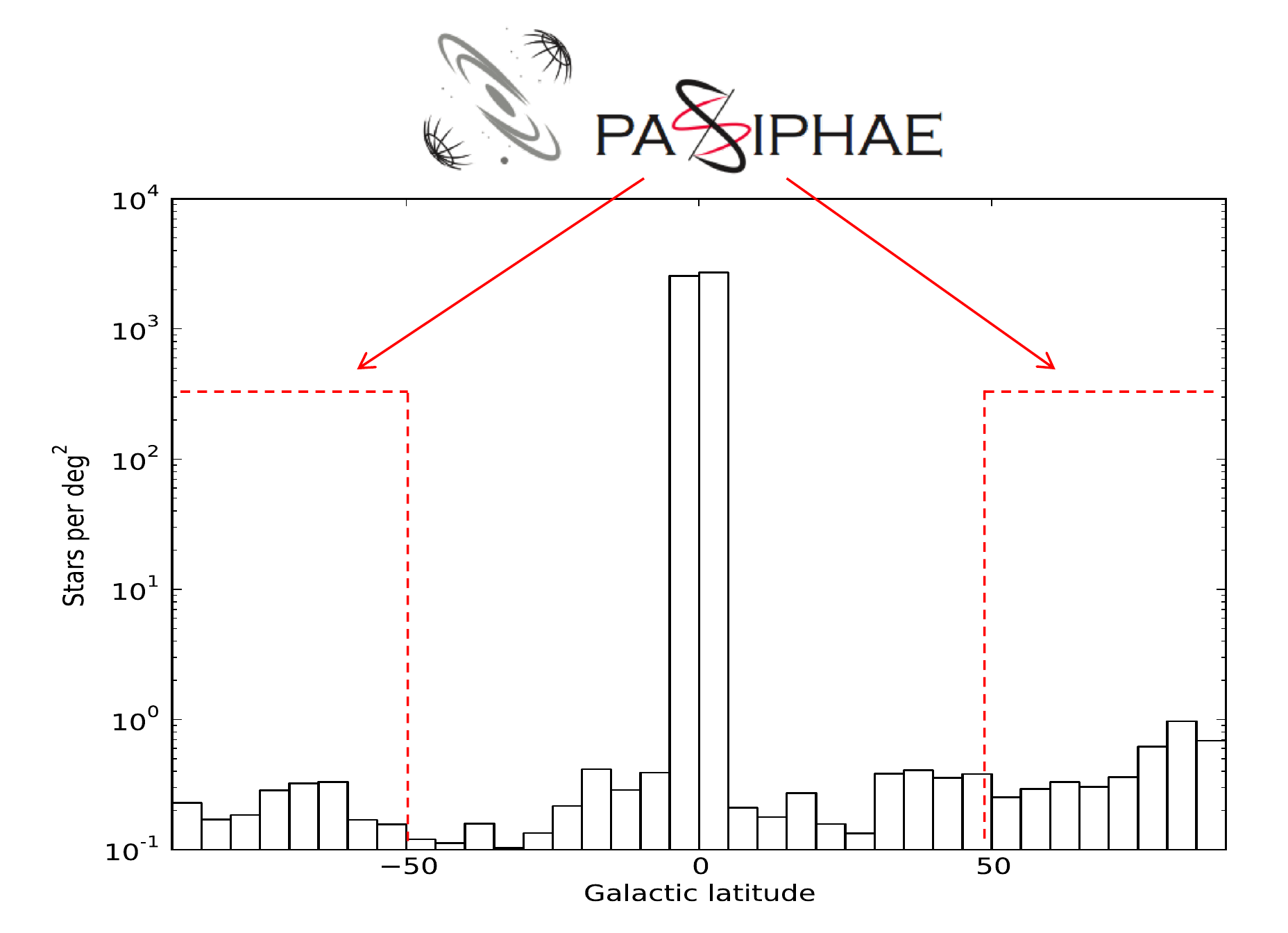}
\caption{Distribution of stars with measured polarization properties as a function of Galactic latitude. Data from NED + GPIPS survey (Clemens et al. 2012). For comparison, the dashed lines mark the lower limit of what PASIPHAE will achieve.
\label{fig:poltodate}}
\end{wrapfigure}

B-mode experiments target high-Galactic latitudes, because of their low dust content. In contrast, surveys of starlight polarization have up to now been driven by interstellar medium science, and as a result they are heavily biased towards the Galactic plane (low Galactic latitudes), where most of the dust lies \citep{2000AJ....119..923H, 2012ApJS..200...21C} (see Fig. \ref{fig:poltodate}). As seen in Figure \ref{fig:poltodate}, at high Galactic latitudes, only very few, bright, (V$<$13 mag) nearby stars have measured polarization properties \citep{2001A&A...368..635B, 2001A&A...372..276B}.

\subsubsection{How polarized will starlight be at high Galactic latitudes?}

Since there is little dust at high Galactic latitudes, the resulting degree of polarization ($p$) will be low. Measurements of starlight polarization of selected bright stars show typical values of 0.1\%. However, these bright stars only sample a fraction of the total sightline \citep[extending out to 600 pc at most,][]{2014A&A...561A..24B}. To obtain plausible lower limits on the expected $p$, we have measured starlight polarization in regions where the dust content is minimal. These pathfinding surveys are deep enough (R$\leqslant$16 mag) to extend out to 2 kpc. We find that the average $p$ induced by the ISM within two of the least dust-extincted regions (0.06 sq. deg) is 0.1\% $-$ 0.2\% \citep{Skalidis}. Although this polarization fraction is too low to be measured in individual stars, optopolarimetric observations of {\em many stars} allows for a much higher sensitivity to the {\em average} polarization fraction of a region. 



\subsubsection{Out to what distance should we measure?}

To achieve the required sensitivity for CMB studies tomographic result, the survey must be able to reach out to the most distant cloud that contributes to the signal detected in dust emission for any given sightline. At high Galactic latitude, the most distant clouds with detectable dust content are the Intermediate Velocity Clouds (IVCs)\footnote{Another class of distant clouds found at high latitude is that of High Velocity Clouds (HVC). These reside at distances of several kpc \citep{Wakker}, where the radiation field (and consequently, dust heating) is weak. Searches for dust emission from these clouds yield non-detections, with the notable exception of a small compact cloud within the HVC Complex M \citep{Peek2009} and possibly a region of a few square degrees in Complex C \citep{Miville}.}, named after their HI emission which is found at velocities inconsistent with simple Galactic rotation \citep{1991A&A...250..499W}. Distance measurements to such clouds place them within 2 kpc of the Sun \citep{Wakker}. Even though IVCs may not cover the entire high-latitude sky, dust-bearing HI gas forms the Galactic HI halo, which extends $\sim 1.6$ kpc vertically from the disk \citep{2011A&A...525A.134M}. Therefore, in order to map the magnetic field out to such distances, our survey must include stars that are more distant than 1.6$-$2 kpc. 


\begin{figure}[ht!]
\centering
\includegraphics[width=0.8\textwidth]{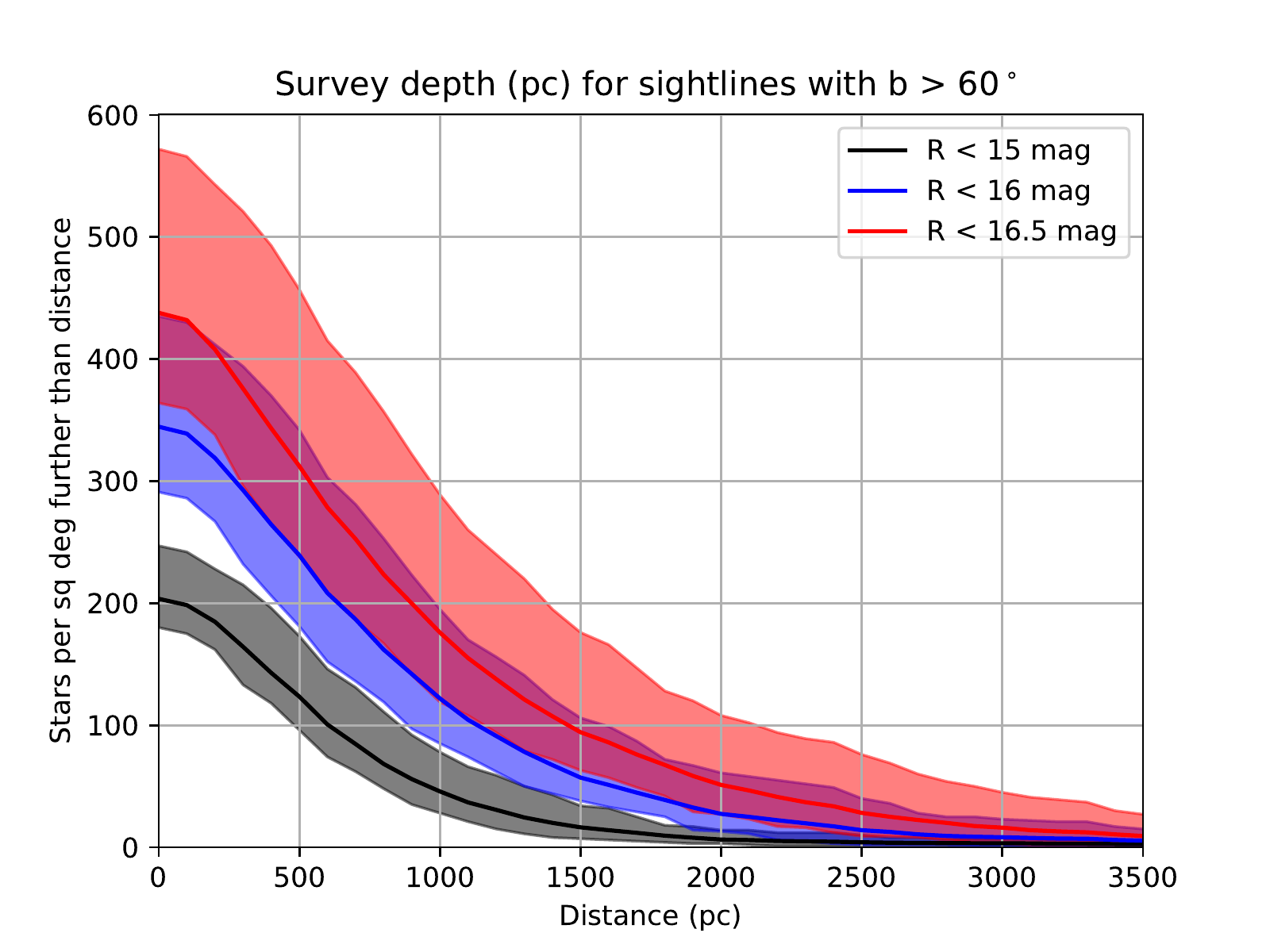}
\caption{Number of stars per square degree that are predicted to lie further than the distance shown in the horizontal axis. For a given magnitude cut in the R band (R $< 15$ mag, black, R $<16$ mag, blue, R $<16.5$ mag, red) shaded areas mark the range of values, while the solid lines show the mean (average over all selected sightlines). The sightlines sample the sky with $b>60^\circ$ uniformly in area. The data are taken from a model of the stellar population in our Galaxy\citep{czekaj} which has been tested against and informed by existing datasets. }
\label{fig:stars_d}
\end{figure}

\subsubsection{How photometricaly deep a survey do we need?}

The distance requirement dictates a strict limiting magnitude of the survey. There should be enough stars per ``pixel" (radio telescope beam) of a B-mode experiment behind the furthest clouds so that a significant statistical average is measured. For most high-latitude sightlines, there are fewer than 5 stars per sq. degree with R $<15$ mag that lie further than 2 kpc (Figure \ref{fig:stars_d}). The density increases to 30 stars per sq. degree for R $<16$ mag and 50 stars per sq. degree for R $<16.5$ mag. Independently of distance, the stellar density at these limiting magnitudes is a few hundred stars per sq. degree. For comparison, currently there is less than 1 star per sq. degree with a measured polarization (Figure \ref{fig:poltodate}) and not always with the required accuracy. 





\section{Beyond the state of the art: PASIPHAE}

If we are to reach our tomographic target in time for the next-generation B-mode experiments, we urgently need a giant leap forward in our capabilities to measure the optical polarization of starlight. To meet this pressing need, we propose a program aimed at performing that necessary leap, by mapping interstellar magnetic fields at high Galactic latitudes in both northern and southern hemispheres, using stellar optical polarization. We call our program PASIPHAE (Polar-Area Stellar Imaging Polarization High-Accuracy Experiment). The main aim of this program is to deliver tomographic cartography of the Galactic magnetic field at high Galactic latitudes.

\subsection{The high-latitude optopolarimetric Survey}\label{survey}

\subsubsection{Two ideal observing sites}

\begin{figure}[ht!]
\includegraphics[width=0.462\textwidth, clip]{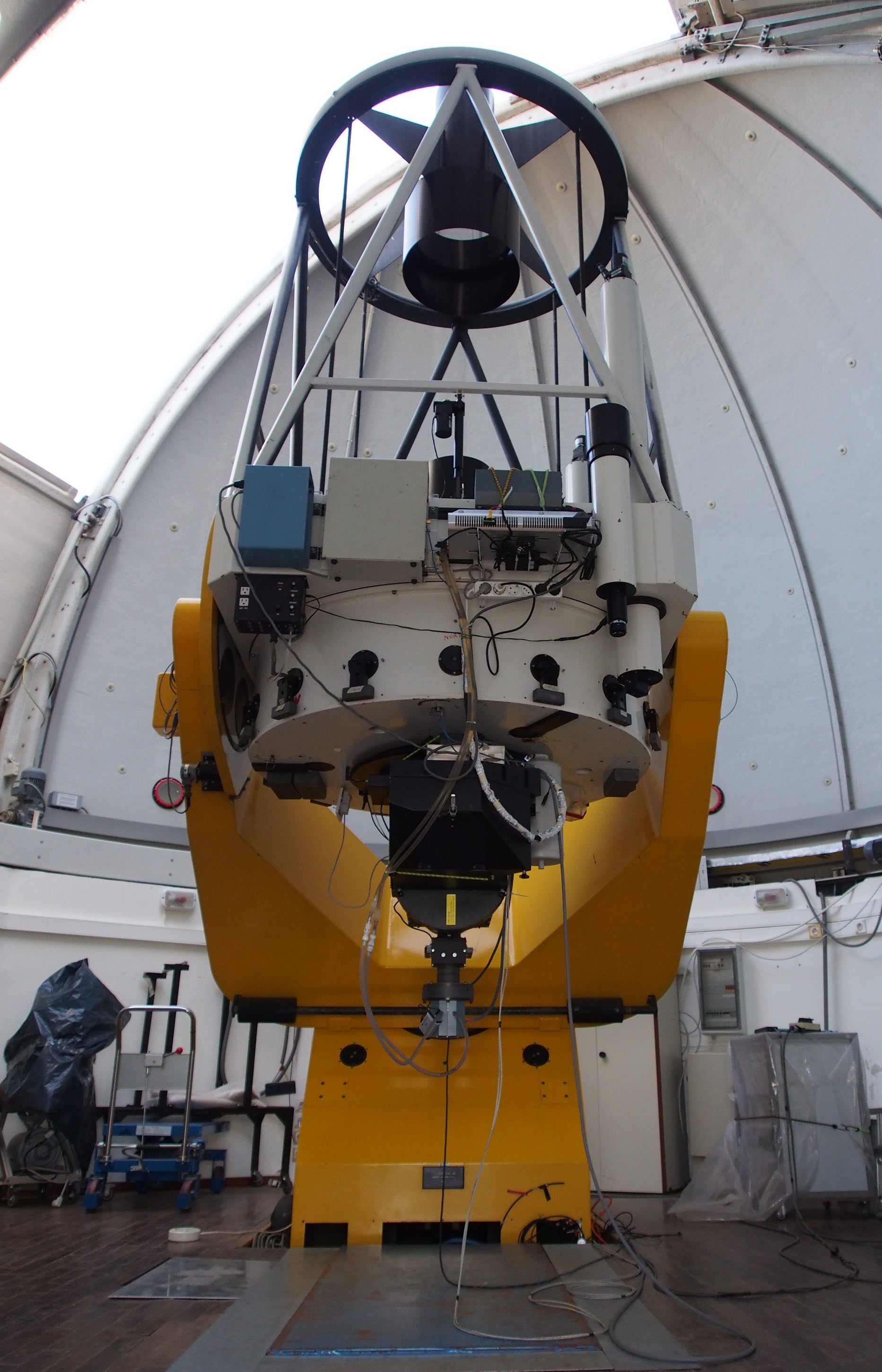}
\includegraphics[width=0.538\textwidth, clip]{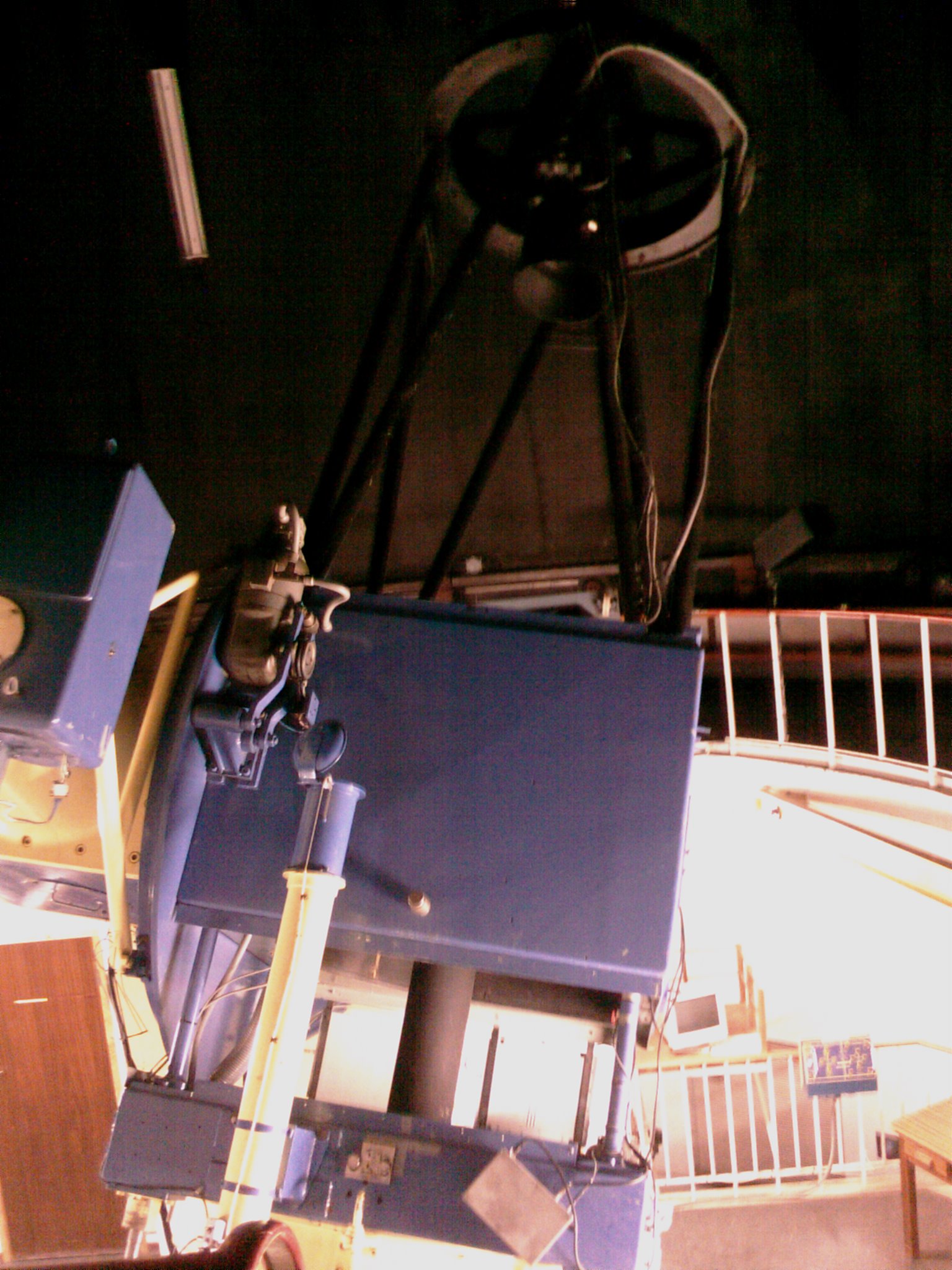}
\caption{Skinakas observatory 1.3 m telescope (left) and SAAO 1 m Elizabeth telescope (right).
\label{fig6}}
\end{figure}

We will operate PASIPHAE simultaneously in the north and south hemispheres, to provide a tomographic map of the Galactic magnetic field in the cleanest (low dust column) areas of both the northern and southern sky.

In the north, the survey will be run at the Skinakas Observatory\footnote{\url{http://www.skinakas.org.gr}}, which is located in the island of Crete, Greece, and it is managed by FORTH\footnote{Foundation for Research and Technology - Hellas \url{https://www.forth.gr}} and the University of Crete. The Observatory is located at a mountain peak of 1750 m and has been operating continuously, hosting the most advanced instrumentation in Greece, since 1986. The site has excellent atmospheric conditions, among the best for astronomical observations in Europe, comparable to several of the largest facilities in the world (typical seeing ~1 arcsec) \citep{2001hell.confE.134B}. The main 1.3 m Ritchey-Chr\'{e}tien telescope at Skinakas (Fig. \ref{fig6}, left) is a uniquely suitable facility for conducting a large area survey. It combines a large field of view $(30^{'} \times 30^{'})$ with a high sensitivity and it can be remotely operated. The Skinakas Observatory has committed 135 nights of observing time per year on the 1.3 m telescope to our program.

In the south, the survey will be run at the South African Astronomical Observatory\footnote{\url{http://www.saao.ac.za}} (SAAO) at Sutherland, South Africa. The Sutherland site, located at 1800 m above sea level, hosts all major telescopes of SAAO. The site features excellent weather and seeing (~1 arcsec\footnote{Erasmus, South African Journal of Science 96, Month 2000}), a semi-desert setting, and absence of light pollution. The 1.0 m telescope at Sutherland (Fig. \ref{fig6}, right) features a somewhat larger field of view $(34^{'} \times 34^{'})$ than the Skinakas 1.3 m, which balances out the difference in diameter in terms of survey speed. The SAAO has committed 150 nights of observing time per year on their fully remotely operable 1.0 m to our program. 

\subsubsection{Survey rate}

PASIPHAE will deliver at least a $3 \sigma$ measurement of a polarization fraction $p \approx 0.5\%$ for a point source of brightness R = 16.4 mag within an exposure time of 15 min at the Skinakas 1.3 m (1.2 m effective) (24 min at the SAAO 1m telescope, 0.94 m effective). With a field of view of 0.25 square degrees (0.35 for SAAO), assuming 8 hours of observing per night and 70\% efficiency, we can survey 8 (7 for SAAO) square degrees per night, corresponding to approximately 1,500 square degrees per year (in both hemispheres). Our survey will cover $\sim$7,500 square degrees over the duration of the PASIPHAE project, which exceeds the sky area from the Galactic pole down to 55 degrees of Galactic latitude in both northern and southern hemispheres.

\subsection{Enabling a revolutionary survey: the WALOP polarimeters}\label{walops}

This survey is enabled by two innovative-technology Wide-Area Linear Optical Polarimeters (WALOPs), designed for high accuracy, sensitivity, and field of view. Their design is based on the highly successful 4-channel design of the RoboPol polarimeter\footnote{\url{http://www.robopol.org}} that has been operating since May of 2013 at the 1.3 m telescope of  Skinakas Observatory. RoboPol has already implemented, in a smaller field of view, some of the technological concepts that will be applied to the entire large field of view of the WALOP polarimeters. RoboPol has been used to conduct extensive path-finding tests for the WALOP polarimeter design and to predict WALOP performance. These tests, as well as preliminary simulations of the WALOP optical and electronics design, confirm that the achievable systematic errors control will be as small as 0.1\% in polarization fraction for stars of $R \lesssim 16$ mag.

\subsubsection{WALOP design principle}

Measurement of partial linear polarization demands multiple measurements of the light from the celestial source through different settings of polarization-sensitive optical components. Conventional polarimeters typically use birefringent beam-splitters and rotating half-wave plates to take the necessary intensity measurements during subsequent exposures of the same source. Their efficiency and accuracy is thus limited by the need for moving parts and multiple, non-simultaneous measurements.

The WALOP polarimeters will instead implement a 4-channel, no-moving-parts, one-shot design, optimized for a wide field survey. This innovative 4-channel design has already been successfully tested with RoboPol \citep{2014MNRAS.442.1706K}. Although RoboPol had in principle wide-field capabilities, it was designed primarily for point source monitoring. For this reason, its sensitivity and accuracy are optimal for the central point source it observes. To minimize cost, all four spots in RoboPol are projected on the same CCD. This however in general reduces the sensitivity of the instrument, by projecting (at low $p$ values) roughly a fourth of the photons of a point source against a full sky (since the sky is an extended source). This problem was solved by placing a mask at the center of the field of view, ensuring that each spot of the central point source is projected on a CCD region against only a quarter of the full sky intensity. However, field stars in RoboPol are measured against the full sky.

The WALOP polarimeters will instead arrange for the four spots to fall on four different detectors, effectively extending the benefits of the RoboPol mask to the entire field. The instrument will thus have features uniquely appropriate for a wide-field survey.
\begin{itemize}
\item {\bf High efficiency}: The 4-channel design ensures that a single exposure per field of view is enough to derive linear polarization properties.
\item {\bf High sensitivity}: The 4-CCD design splits the sky background to four (as is done for the light from point sources), increasing the photometric signal-to-noise ratio over the single-CCD design.
\item {\bf High accuracy}: The 4-CCD design permits the use of fast photometric correction techniques (such as flat-fielding), because light falling on each pixel from either a point source or the sky follows a single optical path through the instrument, which is a necessary condition for flat-fielding. At the same time, having each spot form on a different CCD minimizes blending between spots from different sources.
\item {\bf Large field of view}: $30^{'} \times 30^{'}$ (0.25 square degrees per exposure), enabling a wide-field survey.
\end{itemize}

The WALOPs are currently (Oct. 2018) under construction at the Inter-University Centre for Astronomy and Astrophysics in India, the same optopolarimetry instrumentation lab that developed RoboPol. Commissioning is scheduled for 2019.

\subsection{Limiting magnitude and signal-to-noise ratio} 

We have developed detailed error budgets to estimate instrument performance. The estimates have been validated against the actual performance of RoboPol on the same 1.3 m telescope that will be used for WALOP at Skinakas, and found to be consistent. With these specifications and observing conditions, the WALOPs will be able to deliver a polarimetric accuracy of 0.2\%\footnote{total statistical and systematic: the measurement of a star with polarization degree $p$ will be $p \pm 2 \times 10^{-3}$} for a point source of brightness R = 16.4, within a total exposure time of 15 min at the Skinakas 1.3 m telescope (1.2 m effective), and 24 min at the SAAO Sutherland 1 m telescope (0.94 m effective). 
With 30 stars per square degree further than 2 kpc (even for sightlines with the least stellar density), and an uncertainty on each measurement of $\sim$0.2\% (which includes the systematic uncertainty) we will be able to detect at 3$\sigma$ significance even the furthest clouds with mean $p$ of $\sim 0.1\%$ by averaging measurements within this area (see \citet{Skalidis} for a demonstration of this technique).

\subsection{Tomographic map and CMB foreground applications }

\subsubsection{Magnetic field tomographic cartography}

\begin{wrapfigure}{L}{0.6\textwidth}
\centering
\includegraphics[width=0.59\textwidth, clip]{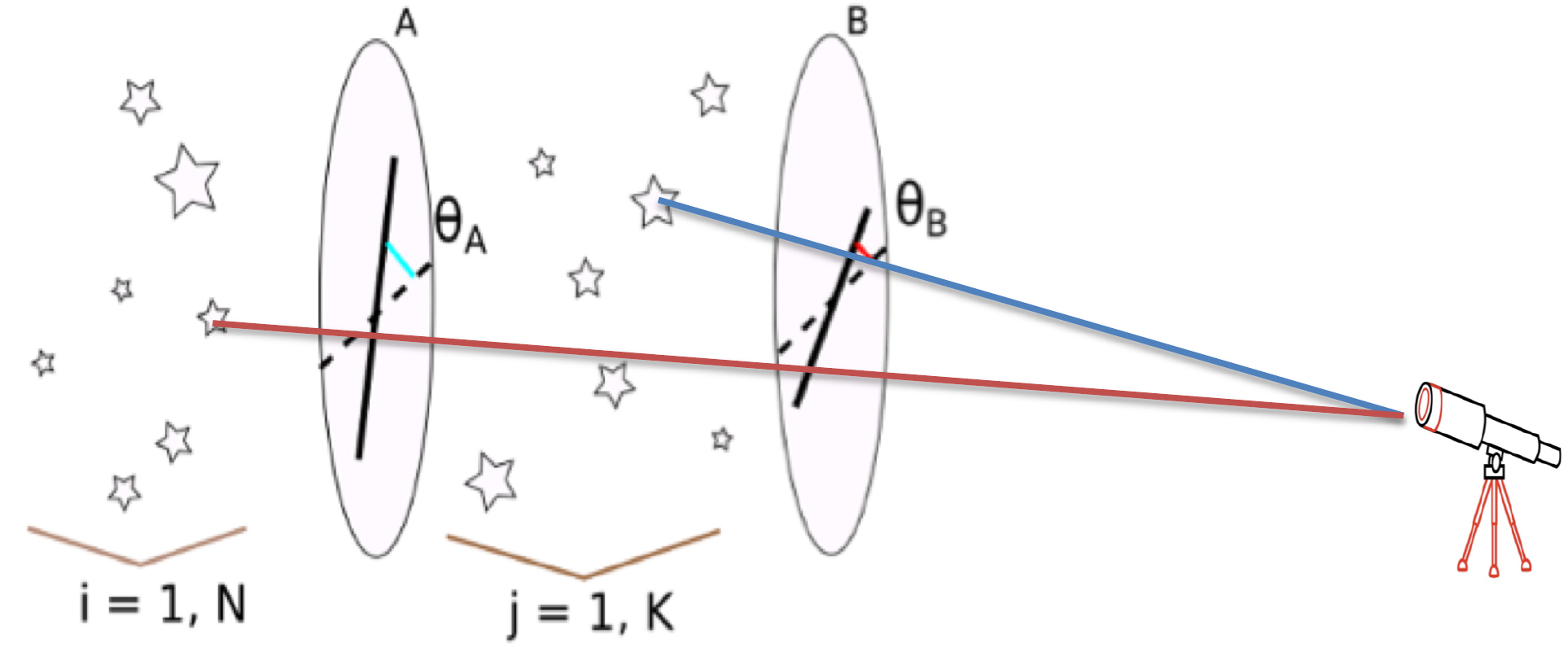}
\caption{Schematic of the distribution of stars with respect to two ISM clouds (with different plane-of-sky magnetic field orientations $\theta_A$, $\theta_B$) lying along a sightline. The clouds are treated as polarizing screens (cloud A on the left, cloud B on the right) inducing polarization on background stars independently from one another. To measure the plane-of-sky magnetic field orientation of cloud A, the effect of cloud B on the polarization of the furthest stars must be removed.}
\label{fig:cartoon3}
\end{wrapfigure}

In order to construct a tomographic map of the Galactic magnetic field, one needs to de-convolve the polarimetric effect of different clouds in between stars with measured properties (see Fig. \ref{fig:cartoon3}). To this end, we will combine our polarization measurements with distance measurements from the on-going Gaia mission.
With this combined dataset we will produce a database of stellar polarization angle, $\theta$, and polarization fraction, $p$, averaged within 3-dimensional, stellar-density-dependent pixels on the sky. These will store the information of how $\theta$ and $p$ vary as functions of distance and sky position. This information will be used to isolate the mean polarization properties of the dusty ISM at different distances. Starting from the nearest stars, the polarizing effect of each portion of the ISM on starlight will be removed incrementally, leading to a final decomposition of the mean plane-of-sky magnetic field orientation at different locations along the line of sight. We present a first demonstration of the technique in an area with two clouds along the line-of-sight using data from RoboPol in \citep{Panopoulou_tomography2018}.
Ultimately, we will provide a database of mean direction of the high-Galactic-latitude magnetic field as a function of location on the sky and distance, including the volume over which the magnetic field is averaged, and its directional uncertainty.




{\bf Masking the CMB B-modes sky }

In \cite{2015MNRAS.451L..90T} we showed that the tomographic effects which limit dust foreground removal in CMB experiments were most severe for lines of sights with clouds with magnetic fields misaligned by more than 60 degrees. A first, immediately achievable goal of the survey is to identify and provide the locations of such sightlines so that they can be excluded from B-mode analyses.

{\bf An improved CMB polarization foreground removal software}

A more sophisticated next step would be to combine our tomographic information with polarized emission data and perform a joint fit of the combined data. We intend to integrate this capability into the COMMANDER software \citep{Eriksen}, a state of the art component separation code that fits parametric models of the foreground emission.

\subsection{Astrophysical applications}

\subsubsection{Structure and dynamics of translucent interstellar clouds }

Diffuse molecular clouds are missed in polarimetric maps close to the Galactic plane, because their signal (either polarized emission from dust or absorption imparted on starlight) is small compared to that of denser clouds along the line of sight. Only in a ``clean" sky region at high Galactic latitudes can the polarization from diffuse clouds dominate. Given that the high Galactic latitudes are, for all practical purposes, unmapped in optical polarization, diffuse clouds are essentially unexplored in absorption.
We will use our tomographic mapping of the interstellar magnetic field to study its morphological structure in diffuse CO-dark molecular clouds. Optopolarimetric observations of starlight polarized through dichroic absorption is the only way to trace the magnetic fields in diffuse clouds, because they are not dense enough to show significant emission in microwave frequencies. Diffuse clouds are best mapped at high Galactic latitudes, because the accumulated ionized carbon emission background from other, denser, more intensely emitting clouds along the line of sight is lower the further away we look from the Galactic plane. For these reasons, even the survey data alone without Gaia parallaxes will represent a giant leap forward in our observational understanding of these clouds, both in terms of number of clouds mapped, as well as in terms of angular resolution of magnetic field mapping. Tomographic mapping of these clouds will provide unprecedented constraints on the molecular cloud formation process that will guide ongoing simulation efforts.
We will use state of the art statistical models to correct for line-of-sight effects in polarimetric observations; to identify and classify structures in molecular clouds; and to connect simulations with observations. In this way, we will produce not only diffuse cloud maps, but also a comprehensive quantitative phenomenological description of the interstellar medium phase straddling
the HI atomic clouds and the dense, star-forming molecular clouds.

\subsubsection{Intrinsic polarization of stars }

If only a single source is observed, intrinsic polarization and interstellar, absorption-induced polarization cannot be distinguished. In a well-sampled wide-field survey, however, sources that are intrinsically polarized stand out, because their polarization fraction and angle are uncorrelated with those of other nearby stars. A high-accuracy, high-sensitivity survey, such as the one we propose, would enable us to not only identify highly polarized sources (like quasars) but also sources that are intrinsically polarized at only a very low polarization level, such as intrinsically polarized stars. Although starlight is unpolarized when it leaves the star, several processes near the star, such as scattering or absorption from dusty material surrounding the star, could result in a low degree of partial linear polarization. Our data would allow, for the first time, to study such cases of intrinsic polarization and seek systematic correlations with stellar properties, such as spectral type and evolutionary stage. Combined with the results of spectroscopic surveys (such as WEAVE\footnote{\url{https://ingconfluence.ing.iac.es:8444/confluence//display/WEAV/The+WEAVE+Project}}) that are currently underway in support of the Gaia mission, this dataset will open new avenues of exploration in stellar astrophysics.

\subsubsection{Unidentified gamma-ray sources }

We will identify candidate optical counterparts for the 100 yet-unidentified gamma-ray sources detected by the Fermi satellite in the PASIPHAE survey area \citep{2015ApJS..218...23A}. The most common emission mechanism for GeV gamma rays from point sources is inverse Compton scattering of soft photons by relativistic electrons. In most cases, these same relativistic electrons also emit significantly polarized Synchrotron radiation at optical wavelengths. Our high-latitude survey is ideally suited to identify such optical counterparts within gamma- ray-source positional uncertainties that will be highly polarized above what is expected from interstellar polarization alone.

\section{Wider Impact}

The study of optical polarization opens a unique, invaluable, and vastly under-unexplored window to the universe. Beyond the search for the inflationary B-modes, a survey of polarization of point sources at optical wavelengths and at high Galactic latitudes would have far-reaching applications in some of the most actively pursued areas in astrophysics and cosmology. Beyond the study of interstellar clouds, circumstellar disks, and high-energy astrophysics (subjects that PASIPHAE collaboration will pursue), astrophysical applications of the survey data include:
\begin{itemize}

\item {\bf Physical models of interstellar dust} Knowledge of both the absorption and emission polarization properties of dust will allow benchmarking, calibration, and improvement of physical dust models [e.g., \citet{2009ApJ...696....1D}]  that are also essential tool in high-accuracy B-mode foreground removal. In addition, this wealth of data will help address open questions in our physical understanding of interstellar dust, including spatial variations in the optical-UV extinction law, peak emission wavelength, and far-infrared opacity per unit of extinction.

\item {\bf Ultra-high-energy cosmic-ray astrophysics} An optopolarimetric survey such as the one proposed here can transform the astrophysics of the highest-energy particles in the Universe. The 3-dimensional map of Galactic magnetic field directions that can be reconstructed through PASIPHAE data will aid efforts to backtrack the paths that cosmic rays of the highest energies (over $10^{18}$ eV) traverse through the Galaxy before reaching us, to improve agreement between their corrected arrival directions and the location of their sources on the sky.

\item {\bf MHD instabilities and the Galactic fountain} MHD instabilities can leave a characteristic imprint on the structure of the Galactic magnetic field. This would have important applications in ISM physics, with prime examples being interstellar turbulence and the origin of the Galactic magnetic field. Similarly, the ``Galactic fountain" model for the gas enrichment of the Galaxy proposes the recycling of hot material back towards the disk along magnetic field lines that run from the center of the Galaxy to the halo and back towards the disk. A detailed mapping of the magnetic field near the Galactic pole region would provide invaluable tests for such models.

\end{itemize}

\subsection{Summary / PASIPHAE science deliverables}
Our program objectives can be summarized as follows:

\begin{enumerate}

\item {\bf High Galactic Latitude Survey}: Combining our novel polarimeters with a massive commitment of observing time by the Skinakas Observatory in Crete, Greece, in the north, and the South African Astronomical Observatory (SAAO) in Sutherland, South Africa, in the south. We will deliver polarization measurements [degree of polarization $p$, and polarization direction $\chi$ (electric vector position angle, EVPA), for surveyed point sources] with a 0.2\% accuracy in degree of polarization of individual stars for over 360 stars per square degree, for over 1500 square degrees per year of survey. 
For the high-Galactic-latitude areas that are the focus of the program, PASIPHAE mapping will take us from the current state of about 1 star per 3 square degree with measured polarization properties, to over 360 stars per square degree. {\em This is a 1000-fold increase for the density of stars with measured polarization.}

\item {\bf Tomographic map and CMB foreground subtraction}: We will combine our survey products with stellar distances from Gaia, and deliver, for the first time ever, a tomographic cartography of the Galactic magnetic field.
\begin{itemize}
\item {\em Sky mask}: We will use this map to identify regions of the sky where clouds along the line of sight have severely misaligned magnetic fields where frequency decorrelation effects are likely to be most severe. We will produce a sky mask for the CMB scientific community to block out these regions in CMB B-mode analyses.
\item {\em Foreground removal software}: In synergy with dust emission data, we will use the survey data to produce an improved model for the polarized emission spectrum of cold dust, which will accurately predict pattern decorrelations between frequencies due to 3-dimensional effects. We will integrate the model in the state of the art CMB foreground-removal software COMMANDER.
\end{itemize}
\item {\bf Astrophysical Applications}: We will use our tomographic map of the Galactic magnetic field and our survey catalogue for an unprecedented and wide-reaching array of astrophysical applications including: studies of the interstellar medium at high Galactic latitudes, focusing on the structure and magnetic field properties of translucent molecular clouds; stellar astrophysics of intrinsically polarized stars, such as systems with circumstellar/debris disks; and high-energy astrophysical sources of cyclotron and/or synchrotron emission, and in particular searches for candidate optical counterparts for the 100 yet-unidentified gamma-ray sources detected by the Fermi satellite in the PASIPHAE survey area.

\end{enumerate}

\section*{Acknowledgements}

The PASIPHAE collaboration acknowledges support from the European Research Council (ERC) under the European Union's Horizon 2020 research and innovation programme under grant agreement No 771282 and No 772253, from the National Science Foundation, under grant number AST-1611547 and the National Research Foundation of South Africa under the National Equipment Programme. 
This project is also funded by an  infrastructure development grant from the Stavros Niarchos Foundation and from the Infosys Foundation.

\end{document}